\documentclass{sigchi}

\toappear{
Permission to make digital or hard copies of all or part of this
work for personal or classroom use is granted without fee provided that 
copies are not made or distributed for profit or commercial advantage and
that copies bear this notice and the full citation on the first page. To
copy otherwise, or republish, to post on servers or to redistribute to 
lists, requires prior specific permission and/or a fee.\\
{\confname{WebSci'12}}, May 2--4, 2013, Paris, France.\\
Copyright 2013 ACM 978-1-4503-1889-1...\$10.00.
}

\usepackage{balance}  % to better equalize the last page
\usepackage{graphics} % for EPS, load graphicx instead
\usepackage{times}    % comment if you want LaTeX's default font
\usepackage{url}      % llt: nicely formatted URLs

\makeatletter
\def\url@leostyle{%
 
\@ifundefined{selectfont}{\def\UrlFont{\sf}}{\def\UrlFont{\small\bf\ttfamily}}}
\makeatother
\def\pprw{8.5in}
\def\pprh{11in}

\usepackage[colorlinks,urlcolor=blue]{hyperref}
\hypersetup{
pdftitle={WebSci Conference Proceedings},
pdfauthor={LaTeX},
pdfkeywords={WebSci, proceedings},
bookmarksnumbered,
pdfstartview={FitH},
colorlinks,
citecolor=black,
filecolor=black,
linkcolor=black,
urlcolor=black,
breaklinks=true,
}

\usepackage{caption}
\usepackage{subcaption}

\begin{document}

\title{Genericity versus expressivity - an exercise in semantic interoperable
research information systems for Web Science}

% \title{Genericity versus expressivity - an exercise in semantic interoperable
% research information systems for Web Science}

% \title{VIVO going Europe - an exercise in semantic interoperable research 
% information systems}

\numberofauthors{3}
\author{
   \alignauthor Christophe Gu\'eret\\
     \affaddr{Data Archiving and Networked Services (DANS)}\\
     \affaddr{Den Haag, The Netherlands}\\
     \email{christophe.gueret@dans.knaw.nl}
   \alignauthor Tamy Chambers\\
     \affaddr{School of Library \& Information Sciences}\\
     \affaddr{Indiana University, USA}\\
     \email{tischt@indiana.edu}
   \alignauthor Linda Reijnhoudt\\
     \affaddr{Data Archiving and Networked Services (DANS)}\\
     \affaddr{Den Haag, The Netherlands}\\
     \email{linda.reijnhoudt@dans.knaw.nl}
   \alignauthor Frank van der Most\\
     \affaddr{Data Archiving and Networked Services (DANS)}\\
     \affaddr{Den Haag, The Netherlands}\\
     \email{frank.vandermost@dans.knaw.nl}
   \alignauthor Andrea Scharnhorst\\
     \affaddr{Data Archiving and Networked Services (DANS)}\\
     \affaddr{Den Haag, The Netherlands}\\
     \email{andrea.scharnhorst@dans.knaw.nl}
}

% Tamy, Linda, Frank, Andrea
%   \alignauthor 1st Author Name\\
%     \affaddr{Affiliation}\\
%     \affaddr{Address}\\
%     \email{e-mail address}\\
%     \affaddr{Optional phone number}
%   \alignauthor 2nd Author Name\\
%     \affaddr{Affiliation}\\
%     \affaddr{Address}\\
%     \email{e-mail address}\\
%     \affaddr{Optional phone number}    
%   \alignauthor 3rd Author Name\\
%     \affaddr{Affiliation}\\
%     \affaddr{Address}\\
%     \email{e-mail address}\\
%     \affaddr{Optional phone number}

% Teaser figure can go here
%\teaser{
%  \centering
%  \includegraphics{Figure1}
%  \caption{Teaser Image}
%  \label{fig:teaser}
%}

\maketitle

\begin{abstract} 
The web does not only enable new forms of science, it also creates new
possibilities to study science and new digital scholarship. This paper brings 
together multiple perspectives: from individual researchers seeking the best
options to display their activities and market their skills on the academic job
market; to academic institutions, national funding agencies, and countries
needing to monitor the science system and account for public money spending. We
also address the research interests aimed at better understanding the
self-organising and complex nature of the science system through researcher
tracing, the identification of the emergence of new fields, and knowledge
discovery using large-data mining and non-linear dynamics. In particular this
paper draws attention to the need for standardisation and data interoperability
in the area of research information as an indispensable pre-condition for any
science modelling. We discuss which levels of complexity are needed to provide a
globally, interoperable, and expressive data infrastructure for research
information. With possible dynamic science model applications in mind, we
introduce the need for a ``middle-range'' level of complexity for data
representation and propose a conceptual model for research data based on a core
international ontology with national and local extensions.
\end{abstract}

% Free keywords 

% Classification in http://www.acm.org/about/class/1998/
%\category{H.1.m}{Information Systems}{Miscellaneous}
\vskip 1em
% To be picked from ACM list of keywords 
% \terms{Ontology; VIVO; RIS}

\section{Introduction}
\label{sec:introduction}
The science of the 21$^{st}$ century, to a large extent is team
science~\cite{borner2010multi}, operating globally, often cross disciplinary, and
fully entangled with the web. The study of science as a 
specific, complex, and social system has been addressed by many research disciplines
for quite some time. The availability of digital traces of scholarly
activities at unknown scale and variety, together with the urgent need to monitor
and control this growing system, is at heart of knowledge 
economies and has brought the question how best to measure, model, and forecast
science back on to the research agenda~\cite{Scharnhorst2012}.

When reviewing the current models of science, it is clear there is no 
consistent framework of science models yet~\cite{boerner2012basicmodeltypes}.
Existing models are often driven by the available data. For example, 
interdisciplinary bibliographic databases (such as
the Web of Science or SCOPUS) use the principle of citation
indexing~\cite{garfield1979citation} from the field of \textit{scientometrics}
to analyse the science system based on formal scholarly communication.
Typical output indicators are counts of publications, citations, and patents.
They form the heart of the current ``measurement of science'' and have been
taken up as data by network science~\cite{bara2002} and Web
Science~\cite{huberman2001}.

This specific kind of output is, however, only a tiny fraction of 
information on science dynamics. Traditionally, the measurement of science
encompasses input indicators (human capital, expenditure), output indicators,
and. where
possible, process information~\cite{godin2005measurement}. Research
Information Systems, around since WWII in Europe, are marking the
shift to ``big science''~\cite{derek1963solla}. However, the input side to
science dynamics, in particular researchers, has been underrepresented in
quantitative science studies for quite some time. This is partly due to the lack
of databases and the problem of author ambiguity in the existing 
database~\cite{scharnhorst2010tracing,2013arXiv1301.5177R}. Information on
researchers has been mainly collected, documented, and curated locally at individual
scientific institutions - and in nation-wide research
information systems, at least in European countries. 

The emergence of the web has transformed this situation completely. The web has
become an important, if not the most important, information source for
researchers and a platform for collaboration~\cite{borgman2007scholarship}. The
extent and diversity of the traces scholars leave on the web has called for
\textit{alt metrics}~\cite{wouters2012users}. It has also triggered the
development of standards and ontologies capable of automatically harvesting this
wealth of information, beyond existing traditional bibliographic reference. 

The wealth of information provided on the web about researcher activities and 
their relations carries the potential for new insights into the global research
landscape. But we are not yet at the point where this data can be both
expressive enough to be useful and easy enough to consume. 

\begin{figure*}[!htbp]
	\centering
	\includegraphics[width=0.9\linewidth]{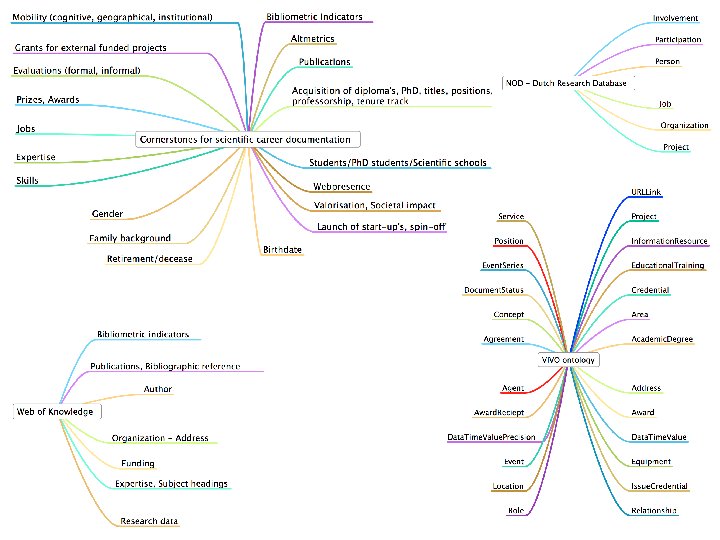}
	\caption{Conceptual space of four different communities dealing with
research information. The variation among these mind maps illustrates
the difference in size, granularity, scope, and expression of the different
information systems with which they are associated.}
	\label{fig:webscience}
\end{figure*}

To illustrate the current situation we display the conceptual space of 
communities dealing with research information in form of four mind maps
(\textit{c.f.} Figure~\ref{fig:webscience}). In the upper left corner we brought
together concepts, which are relevant from the perspective of scientific career
research and often conducted qualitatively, with rich factual evidence, which is hardly
interoperable or scalable. For this mind node we drew on current discussions
and first results~\cite{most2012} in a FP7 framework programme ACUMEN, Academic
Careers understood by Measurements and Norms
(see \url{http://research-acumen.eu/}), where sociologists and scientometricians
work together. In the right lower corner we display the main classes of an
ontology for research information (VIVO\footnote{\url{http://www.vivoweg.org}})
developed in the US. In the upper right corner, the main tables of a Dutch
Research Information Database (NOD-NARCIS) are displayed, and in the lower left
corner is a selection of information and concepts which can be retrieved
using different fields in one of the leading cross-disciplinary bibliographic
databases - the Web of Knowledge. Although, the mind map sketches are different
in nature, from formal schemes to collection of aspects, this
illustration shows their difference in size, granularity, scope, and expression or
semantics.

% \subsection{What to expect from this paper?}
% In the following we .... 

In this work we argue for the need of a scalable, interoperable, and
multi-layered data representation model for research information system (RIS).
Science of science and modeling of science dynamics raise and fall with a
consistent measurement system for the sciences. The contributions of this paper
are as follows:
 \begin{itemize}
 \item A highlight of information loss happening when expressing data with
 generic ontologies;
 \item The introductions of the notion of levels of semantic agreement for
 expressing research data;
 \item A multi-layered ontology based on the above definition.
\end{itemize}
% 
% 
% Two perspectives are brought together in this paper: 
% \begin{itemize}
% \item the perspective of the study of scientific careers via data published on
% the Web and Web technology can empower individual researchers to communicate
% about their activity; 
% \item the need of institutional, national and international monitoring tools for
% research capacities including human capita research questions of an appropriate
% semantic data representation.
% \end{itemize}
% 
% Those perspective are discussed in the context of a very concrete and defined 
% task: the expression of a set of records of researchers at the VU exported from
% the NARCIS database with the study of the relations it contains as the final
% goal. This work started off as an exercice in expressing this data using the
% VIVO ontology and was later extended towards the definition of a multi-layered
% semantic approach.
% 
% The contributions of this paper are as follows:
% \begin{itemize}
% \item An highlight of information loss happening when expressing data with
% generic ontologies;
% \item The introductions of the notion of levels of semantic agreement for
% expressing research data;
% \item A multi-layered ontology based on the above definition.
% \end{itemize}

The remainder of the paper describes the landscape of
research data publication before diving into the details of a specific Dutch
case. We thereafter introduce our proposed multi-layer conceptual model for a
research ontology and conclude in its potential for documenting research.

%According to EuroCRIS\footnote{\url{http://www.eurocris.org}, Current
%Research Information Systems - a network in Europe} research information
%encompasses information on: people (including their positions), organizations,
%projects, funding programs, grants, courses and research
%results~\cite{sheppard2010learning}. 
\vskip 4em

\section{Current landscape of RIS}
\label{sec:current}

\subsection{Publishing research data}
In order to publish re-usable research data, one has to think in terms of
standards and publication media. While the web imposes itself as the
publication platform, the question of standards remains open and has been long
investigated.

First efforts in standardisation have been undertaken from the traditional
research information communities. One example is the ``CERIF'' standard
developed by EuroCRIS\footnote{\url{http://www.eurocris.org}}. This standard
defines a set of generic classes and properties used to describe research data.
The serialisation format used for the data is XML, although an RDF version is
being considered\footnote{\url{http://spi-fm.uca.es/neologism/cerif}}. The
content management system (CMS) ``METIS'', popular in the Netherlands, uses this
standard to store and expose research data. This standard has also been used
for the Dutch portal ``NARCIS''\footnote{\url{http://www.narcis.nl/}}.

The Web of Linked Data is a way of combining the publication platform and the
standards. More recent efforts have been made in this direction via
a number of ontologies and publication platforms. The initiative
LinkedUniversities\footnote{\url{http://linkeduniversities.org/lu/}} provides a
reference towards these systems and highlights their practical use.
VIVO a United States based open source semantic web application is another such a system.
The application both describes and publishes data, using RDF to encode the data and
OWL for the logical structure.In addition to its own classes and properties, 
the VIVO ontology incorpates other standard ontolgies thus increasing its 
interoperability~\cite{rickart2012}. However, the ontology relies heavly on the 
US academic model which limits its ability 
to accurately represent researchers in other systems. 

%Tamy could add a paragraph here; and add why it is so appealing - the  designed
%front it great and you easily see what went wrong and what is ok, please do
%refer to the adoption of the system on the monitoring website of cns, and put
%in a link to the ontologies which all already went into VIVO (the link you
%shared with us)

VIVO and CERIF based CMS have been successfully put in use at many institutions.
Still, the landscape of research information is very scattered and far from
being connected. One of the reasons for this is a lack of agreement upon
semantics for the data. Efforts have been made to align VIVO
and CERIF~\cite{Lezcano2012} but the main problem remains that data publishers
essentially have to choose between using a globally agreed upon representation,
which is less expressive as a result of covering a vast amount of heterogeneous
information (CERIF), or a very expressive and specialised ontology (VIVO), which
is difficult to map to other ontologies of similar complexity.

\subsection{The Dutch case}
In the Netherlands, we find the following situation. All 13 universities (14
with the Open University) use a system called METIS to register and document
their research information~\cite{dupuis2011}. In practice, information is
usually entered in METIS centrally by a person in the administration although,
sometimes individual accounts to METIS are created. Aside from those unconnected
local implementations of one system, higher education in the
Netherlands embraced the Open Access Movement with a project called DARE.  This lead
to an open repository for scientific publications. Moreover, a web portal
to Dutch research information exists - NARCIS - which harvests publications from
open repositories, but also entails a very well curated (and still manually
edited) research information database (NOD) with information about the scientific staff
of about 400 university and outside university research
institutions~\cite{dijk2010, reijnhoudt2012}.

As Oskam and other Dutch researchers already pointed out in 2006, ``the researcher
is key''~\cite{oskam2006harvex}. Outside of institutional RIS this idea is
prolific in Web 2.0. platforms such as Mendeley and Academia.edu. They have been
designed around the needs of scholars. General social network sites
such as LinkedIn - which is very popular for professionals in the Netherlands -
and Facebook also profile themselves as outlets for individual researchers. This 
leads to a situation where user-content driven systems compete for the limited
time and resources of an individual researcher and where, as a result, snippets
of the oeuvre and academic journey of a researcher can be found at different
places, recorded in different standards, and with different accuracy. The
question raised in the 2006 paper: ``How can we make the CRIS\footnote{CRIS
stands for ``Current Research Information System''} a valuable and attractive
(career) tool for the researcher?''~\cite[p. 168]{oskam2006harvex} is still
waiting to be answered in a standardized way. 

The purpose of documentation of science (and of careers of researchers) has
grown far beyond the effective information exchange. Research evaluation relies
heavily on indicators computed (semi) automatically from databases and the web. 
Currently, individual careers of researchers are very much influenced by 
indicators which are built on activities for which large amounts of standardised
data are available. Prominent examples are journal impact factor or the H index.
But, a researcher is not just a ``paper publication machine''. Grant acquisition
is another important ``currency'' in the academic market - for individuals on
the job market, as well as, for institutions competing for funding. Teaching is an
area which is monitored locally and institutionally, but for which no
cross-institutional databases exist.
Moreover, researchers are no longer loyal to one institution, one country, or one
discipline for their whole life. There is an increasing need for
cross-discipline and cross-institutional mapping of whole careers.

%at least I'm not aware of them, so maybe a more careful formulation would be better
% Projects as ACUMEN~\cite{acumenproject} explore ways to empower the individual 
% researcher and to develop guidelines how best to present your academic profile
% to the outside world. From the perspective of careers the following elements
% seem to be in particular relevant...

%Frank could add a paragraph here
%This significant amount of efforts, and research projects such as ACUMEN, 
%highlight an interest in sharing information research among institutions. 

\subsection{Tracing scientific careers}
Projects such as ACUMEN look into current practices of evaluation and peer
review to empower the individual researcher and develop guidelines for how best
to present your academic profile to the outside world. ``ACUMEN'' is the acronym
for Academic Careers Understood through MEasurements and Norms. In this project,
we analyse the use of a wide range of indicators - ranging from traditional
bibliometrics to alt-metrics and metrics based on Web 2.0 - for the evaluation
of the work of individual academics. One of the author of the present work,
Frank van der Most, also conducted interviews to investigate the impact or
influence of evaluations on individual careers.  For his work the
following events are of interesting in tracking an academic career:
\begin{itemize}
\item Birth of the academic;

\item Acquisition of diploma's and titles, in particular MA diplomas (and
equivalents), PhD/Dr. diplomas, habilitiation, professorships of sorts and
levels;

\item Jobs, in universities and academic research institutes, but also in
non-academic organisations. The latter is interesting because people move in,
out, and sometimes back into academia;

\item Particular functions within or as part of the job(s): director of studies
(teaching), research-coordinator, head of department, dean, vice dean (for
research, education, or other), vice-chancellor/rector, board member of
faculty/school/university/institute;

\item Launch of start-ups/spin-outs or people's own companies. It could simply
be a form of employment, but start-ups or own companies may indicate economic or
other societal value of academic work;

\item Prizes;

\item Retirement and decease.
\end{itemize}

For the study of the impact, or influence, of evaluations an overview of
someone's career is necessary to ``locate'' influential evaluations. This
``location'' has multiple dimensions. One is the calendar time, \textit{i.e.} on
which date or in which year did an influential evaluation take place. Based on
time, geographic, and institutional location the context of a particular
evaluation event can be reconstructed. Scientific careers follow patterns which
are influenced by current regimes of science dynamics (including evaluations).

% Together
% with geographical and institutional location, this places the particular
% evaluation in a context in which particular career regimes and evaluation
% regimes exist.

Another important dimension concerns the location of an evaluation (or
any event) within someone's career. If two academics apply for the same job, the
location in time and place is the same, but if one is an early-career researcher
and the other is halfway through his/her career, this clearly makes a large difference to
how their applications are being evaluated and how the evaluation
results are likely to impact their respective careers. A rejection may have a
bigger impact on the early-career researcher than on the mid-career researcher.

Another ACUMEN sub-project investigates gender effects of evaluations and
includes an analysis of performance indicators on research careers. This is
planned to be a statistical analysis which would require some form of career
descriptions.

One of ACUMEN's central aims is to identify and investigate bibliometric
indicators that can be used in the evaluation of the work of individual
researchers. A major point discussed in the ACUMEN workshops is the realisation
that researchers have a career or a life-cycle which contextualises the values
of bibliometric indicators.

Although the events listed above are interesting for ACUMEN, these events, or a
sub-set or extension thereof, is likely to be interesting to many career
studies. For example, productivity-studies would relate academic production of
texts~\cite{Carayol2006,Falagas2008,Levin1991}, courses taught, and other outputs
to someone's career stage or career paths. An academic's epistemic development
(their research agenda) could be studied in relation to career
stages~\cite{Horlings2012} or mobility.

To be able to trace the co-evolution of individual career paths and the social
process of science for larger part of science, one would need a different kind
of information depending on the study being undertaken. 

% While reporting about success and failure,
% discussion and remaining ambiguities of this exercise we touch upon more general
% questions such as ....

%OLD: However, individual researchers provide information to very different 
%sources such as .... 
%Christophe: here your thoughts from the symposium
 
% However standardisation of this data is subject to an interoperability vs 
% expressivity trade off. In this paper we report about mapping the VIVO ontology
% to an ontology used in NARCIS. We discuss the ambiguity of mapping using
% different specific cases. We argue for the need of a more generic core
% vocabulary, and develop workflows how to active this 
% ANDREA: the last half sentence is a bit of a place holder

% Cite http://www.dotac.info/ as related work\input{problem}

\section{Towards a core research vocabulary}
\label{sec:approach}
The challenge when designing a standard for sharing data is to make it generic
enough so that aggregation makes sense, while being specific enough so 
institutions can express the data they need.

As it is highlighted by the two most popular search tools, consuming data
exposed via VIVO from a number of external
sources\footnote{See \url{http://nrn.cns.iu.edu/} and
\url{http://beta.vivosearch.org/}} at the international level, only the most
general concepts such as ``People'' make sense. On the opposite, the search
features offered by a national portal such as NARCIS proposes a number of
refined search criteria. These two extremes of the data mash-up scale show that
depending on the study being done, different levels of semantics agreement are
likely to be put into use.

In contrary to XML schemas, Semantic Web technologies make it possible to
express data using an highly specified model while also making it available
using a more general model. The technology of particular importance here is
``reasoning'', that is the entailment of other factual valid information from
the facts already contained in the knowledge base. For instance, if an RDF
knowledge base contains a fact assessing that ``A is a \textit{researcher}'' and
another stating that ``Every \textit{researcher} is a \textit{person}'', the
system will infer that ``A is a \textit{person}''.

\begin{figure}[!htbp]
\centering
\includegraphics[width=0.9\linewidth]{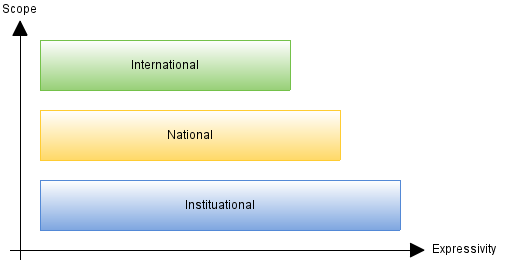}
\caption{The proposed model of multi-layer ontology and its trade-off between
scope and expressivity. At the lowest level, institutionaly defined semantics
have the highest expressivity but the lowest scope.}
\label{fig:layers}
\end{figure}

Leveraging this, it is possible to extend ontologies by refining the definition
of classes and properties. The most refined versions of the concepts will
inherit from their parents. We argue that for research information systems,
three levels are necessary (see Figure~\ref{fig:layers}). First, an
international level containing a set of core concepts that can be used to build
data mash-up on an international scale. Then, a national level extending the
previous core level with concepts commonly agreed upon nation wide
(\textit{e.g.} positions). Last, an institutional level where every institution
is free to further refine the previous level with its own concepts and
properties that matter to its network.

As a feasibility assessment and to propose a first model, we hereafter
introduce a core ontology and two national extensions. This proposal is based on
related work, existing ontologies, and our personal experience but stands more
as a first iteration of work in progress rather than a definitive model.

% It might be too complicated to deal with as such a note saying why we are
% focusing on position is might be appropriate.  Specifically that it is where
% there is the largest variance between countries. 

\subsection{Conceptual models}
Conceptual models allows for the representation of classes and properties of a
knowledge base, along with their relations, in an abstracted way. The proposed
conceptual models that we hereafter introduce are not dependent on the
technical solution implementing them. There is however, as highlighted
previously, an advantage in using Semantic Web technologies for this. This
point is discussed in details in the following, after the introduction and the
description of the three proposed conceptual models.	

\subsubsection{Core model}
\begin{figure*}[!htbp]
	\centering
	\includegraphics[width=0.9\linewidth]{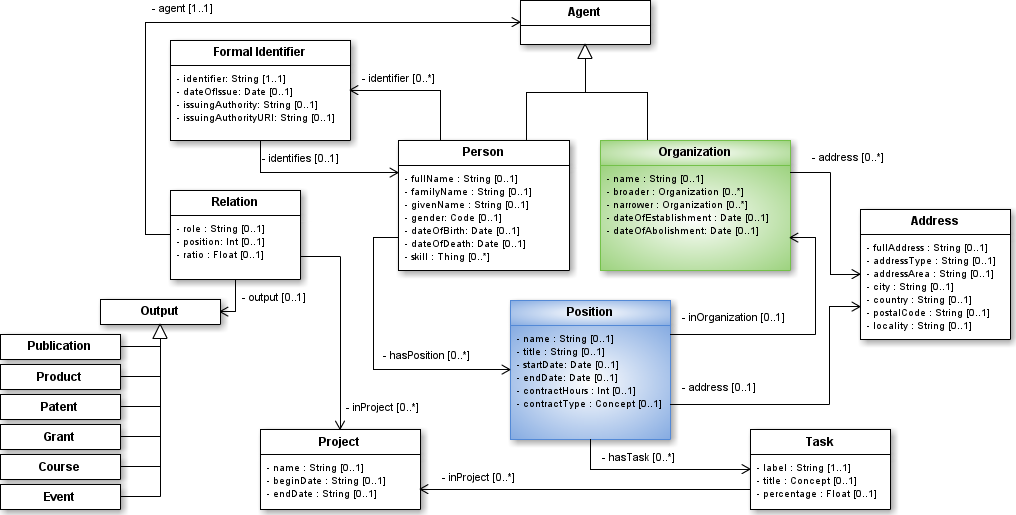}
	\caption{Conceptual model of the core ontology. This model describes the
	minimum set of classes, relationships, and properties needed to describe a natural
	person and trace his scientific career. These classes can be further extended
	by national and local ontologies to account for specificity. As an
	example, the coloured classes are extended in two national ontologies in
	Figure~\ref{fig:extensions}}
	\label{fig:core}
\end{figure*}

The model depicted in Figure~\ref{fig:core} is a proposal for a core research
ontology based on the work being done on CERIF, the VIVO ontology, the Core
vocabularies~\cite{CoreVocab}, and the data needs of ACUMEN. As part of its goal
to study the scientific career through the research data made available, ACUMEN
needs a number of information related to individuals, such as but not limited
to:
\begin{itemize}
	\item Grants/project applications - both applied and granted. This in
	relation to persons (applicants of various sorts) and organisations
	(applying/receiving institutes, main and sub-contractors, funding
	institutes);

	\item Skills. For instance, ``Leadership'' or ``Artificial Intelligence''.
	There is no limit to the definition and several thesaurus could be implied;

	\item Networks or network relations. Relation between persons and
	organisations, but also between persons and results are of particular
	importance;

	\item Memberships of scientific associations or academies;

	\item Conferences visited or organised. 
\end{itemize}

The model contains classes to define individuals, projects, scientific output,
positions and tasks. A generic ``Relation'' can be established between authors
and papers, or teachers and courses taught. The exact meaning of the relation
is to be defined either by sub-classes of it or by using the property ``role''.

\subsubsection{National extensions}
The second level of semantic agreement is that of national extensions. Based on
the core concepts, these extensions allows for the modeling of concepts actually used
in the country - using the language and terminology of that country. When
building such an extension, the main assumption made is that there is a level of
agreement that can be reached on a national basis. 

\begin{figure*}[!htbp]
\centering
\begin{subfigure}[b]{\textwidth}
  \centering
  \includegraphics[width=0.8\linewidth]{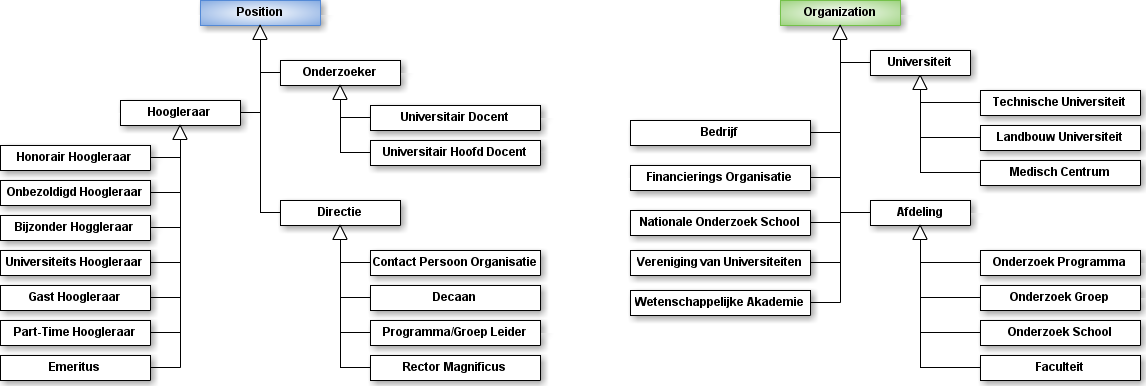}
  \caption{Conceptual model of the extension for the Netherlands}
  \label{fig:dutchext}
\end{subfigure}
\vskip 1em

\begin{subfigure}[b]{\textwidth}
  \centering
  \includegraphics[width=0.8\linewidth]{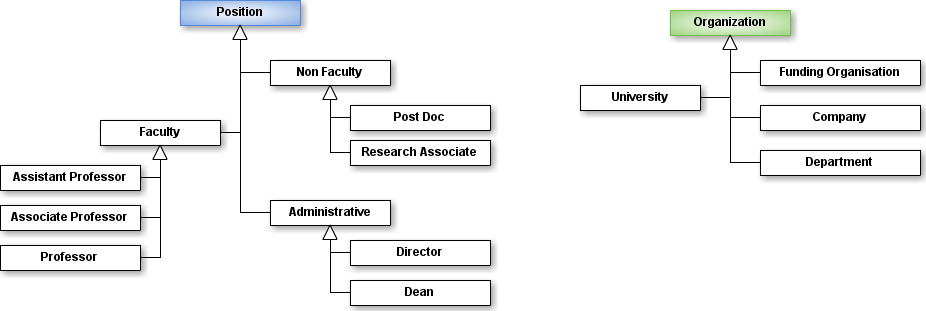}
  \caption{Conceptual model of the extension for the US}
  \label{fig:usext}
\end{subfigure}

\caption{Example of two national extensions of the core model. These extensions
allow for expressing the particularities found in the national system while
grounding their semantic on the more generic concepts.}
\label{fig:extensions}
\end{figure*}

An example of national extension is given in Figure~\ref{fig:extensions}. This
extension extends the core ``Position'' and ``Organization'' classes to define
the type of positions and organisation commonly found in the Netherlands
(Figure~\ref{fig:dutchext}) and the US (Figure~\ref{fig:usext}). The classes
depicted in the Dutch extension are those found in NARCIS, and as
such represent the union set of all the specific classes used within the
research institutions in the Netherlands\footnote{We must
note here that this classes are not defined by an authority but are
rather crowd-sourced. A more accurate, authoritative, list would have to be
defined by an national entity.}.

It can be observed that the Dutch extensions shows a high level of variety, with
some classes that could be replaced with other model mechanisms, such as the
``part time Hoogleraar'' class which is actually a ``Hoogleraar'' contracted
with less hours.

We also note from Figure~\ref{fig:usext} that the national level has to be kept
generic in the US because of the variation observed locally. In the US, many titles
and/or positions are essentially at the discretion of the individual institutions
(with some direction from the American Association of University Professors (AAUP)), 
thus a very detailed national ontology is not appropriate. However, for countries with a more centralised
model and using title and positions officially described, more detail can be
added at this level thus increasing semantic understanding.  The national level
allows for this grey area adaption instead of the current two level ``very
general'' to ``very specific'' model.

\subsubsection{Local extensions}
Local extensions are the most specific level of specification we propose for
this approach. These can be used to specify concepts and relations that are
understood within a given sub community inside a country. For instance, in
the Netherlands, the research institution KNAW defines an additional position
``AkademieHoogleraar'' for ``Hoogleraar'' which are appointed to universities
but directly affiliated to KNAW. This additional position is only used by
some institutions and for this academy - here, the ``Akademie'' in
``AkademieHoogleraar'' implicitly refers to KNAW.

% Obs: both have universities, with same semantics so we could move that up.

\subsection{Implementation}
Prior to its concrete use, the proposed conceptual models have to be turned
into an RDF based vocabulary. This vocabulary also has to be hosted under a
domain name.

\subsubsection{Vocabulary terms}
There are a large number of vocabularies published on the Web. The proposed
models can effectively leverage most of their properties and classes from one
of these existing sources of terms, having fewer new terms to introduce. In
particular, the following vocabularies are to be considered:
\begin{itemize}
	\item FOAF\footnote{\url{http://xmlns.com/foaf/spec/}}, for the description
of the persons;
	\item BIBO\footnote{\url{http://bibliontology.com/}}, for the publications;
	\item LODE\footnote{\url{http://linkedevents.org/ontology/}}, for the
description of events;
	\item SKOS\footnote{\url{http://www.w3.org/2004/02/skos/}}, for the
description of thesaurus terms such as those used to describe researchers'
skills;
	\item PROV-O\footnote{\url{http://www.w3.org/TR/prov-o/}}, to add
additional provenance information to the data being served.
\end{itemize}

We also note that, by design, there is a significant overlap between the
conceptual model of Figure~\ref{fig:core} and that defined in the Core
Vocabularies for Person, Location and registered Organisations in~\cite[page
10]{CoreVocab}. This allows for the proposed core vocabulary for research to be
defined based on these other core vocabularies defined by JoinUp and formalised
by the W3C in the context of the Working Group on Governmental Linked Data
(GLD)~\footnote{\url{http://www.w3.org/2011/gld/wiki/Main_Page}}.

\subsubsection{Ontology hosting}
The domain name at which an ontology is being served is, as for the data
itself, often seen as indication of the person, or entity, in charge of
supporting the ontology. To account for this, we envision the hosting of the
core ontology and its extensions done at institutions matching the scope of the
level of agreement. That is, an international organisation for the
international layer, a national organisation for the national layer, and the
institutions themselves for the local extensions. More concretely, such an
hosting plan could be materialised as having: the core ontology being served by
the W3C, the Dutch national ontology by the VSNU\footnote{the association in
charge of the collective labour agreement for Dutch universities and other
cross-institution regulations on salaries and positions}, and the local
extension from the KNAW by the KNAW.

% Note: look at http://www.eurocris.org/Index.php?page=homepage&t=1
% Note: the core vocabulary could eventually being served by the W3C as some other core vocabularies, see http://ec.europa.eu/isa/
% Note: look at http://www.rechercheisidore.fr/

\section{Conclusion}
\label{sec:conclusion}
This paper operates at different levels. At the core it proposes a model to
semantically describe data in Research Information Systems in a way which
allows to aggregate but also to deconstruct if needed. It does so based on
experiences with standards and data representation in the past and looking into
very concrete practices - taking a VIVO implementation exercise in the
Netherlands as point of reference and departure. 
A next shell of considerations around those specific mappings is added when we
incorporate research outside of the traditional area of scientific information
and documentation. Science and technology studies, science of science, and 
scientometrics have produced over decades of insights in the structure and dynamics
of the science system. 
A wealth of information is available in this area, most of it case-based
evidence. We include the aims and achievements of an on-going EU FP7 funded
project (ACUMEN) which, in itself tries to combine bibliometric and
indicator-based research with interviews, survey, and literature studies. The
target subject of this project is the researcher. 
It is also the researcher which is targeted by Research Information Systems, 
and it is the researcher which is the innovative driver for science dynamics.
Bibliometric indicators are heavily based on standards, part of them shared with RIS.
What makes the ACUMEN project and the perspective of scientific career research
so interesting for the design of future research information systems is
the identification of factors relevant for career development which are not yet
covered by current standards, databases, or ontologies.
The last and most visionary shell in this paper is to design research
information systems which can be used for science modeling.
In the general framework developed by Borner et al. science models can be
developed at different scales of the science system, from the individual
research up to the global science system; they can differ in geographic coverage,
as well as, in scales of time. In any case, the ideal would be having one data
representation which can be scaled up and down along those different dimensions,
and not singular data samples in incomparable measurement units not relatable for particular 
areas of the dynamics of science.
Our main argument is to provide a data representation which is retraceable - if
needed - towards its specific roots and at the same time can be aggregated. In
such a ``measurement system'' we would find a middle layer of data
granularity on which basis complex, non-linear models can be validated and
implemented, to better monitor and understand the science system.    
%cite her paper in the book here
 
\section{Acknowledgments}
This work has been supported by the ACUMEN project FP7 framework. We would like
to think our colleagues Ying Ding,
Katy Borner, and Chris Baars for their comments and support during this work.
% Balancing columns in a ref list is a bit of a pain because you
% either use a hack like flushend or balance, or manually insert
% a column break.  http://www.tex.ac.uk/cgi-bin/texfaq2html?label=balance
% multicols doesn't work because we're already in two-column mode,
% and flushend isn't awesome, so I choose balance.  See this
% for more info: http://cs.brown.edu/system/software/latex/doc/balance.pdf
%
% Note that in a perfect world balance wants to be in the first
% column of the last page.
%
% If balance doesn't work for you, you can remove that and
% hard-code a column break into the bbl file right before you
% submit:
%
% http://stackoverflow.com/questions/2149854/how-to-manually-equalize-columns-
% in-an-ieee-paper-if-using-bibtex
%
% Or, just remove \balance and give up on balancing the last page.
%
\balance

% If you want to use smaller typesetting for the reference list,
% uncomment the following line:


\small
\begin{thebibliography}{10}

\bibitem{VIVOWeb}
{VIVO} {W}eb.
\newblock \url{http://vivoweb.org}.

\bibitem{Core}
e-{G}overnment {C}ore {V}ocabularies: {T}he {SEMIC.EU} approach, 2011.
\newblock Retrieved from European Commission: \url{
  http://joinup.ec.europa.eu/sites/default/files/egovernment-core-vocabularies.p
  df}.

\bibitem{CERIFS}
Cerif -- {1.3} {S}emantics: Research vocabulary, 2012.
\newblock Retrieved from \url{
  http://www.eurocris.org/Uploads/Web%20pages/CERIF-1.3/Specifications/CERIF1.3_Se
  mantics.pdf}.

\bibitem{CoreVocab}
Core vocabularies specification, 2012.
\newblock Retrieved from European Commission: \url{
  https://joinup.ec.europa.eu/sites/default/files/Core_Vocabularies-Business_Loca
  tion_Person-Specification-v0.2_1.pdf}.

\bibitem{bara2002}
Barab{\'a}si, A.
\newblock {\em {Linked: The New Science of Networks}}.
\newblock Cambridge, Mass.: Perseus Publishing, 2002.

\bibitem{borgman2007scholarship}
Borgman, C.
\newblock {\em Scholarship in the digital age: Information, infrastructure, and
  the Internet}.
\newblock MIT press, 2007.

\bibitem{boerner2012basicmodeltypes}
B{\"o}rner, K., Boyack, K., Milojevi{\'c}, S., and Morris, S.
\newblock An introduction to modeling science: Basic model types, key
  definitions, and a general framework for the comparison of process models.
\newblock In {\em {Models of Science Dynamics}}, A.~{Scharnhorst},
  K.~B{\"o}rner, and P.~{Van den Besselaar}, Eds., Understanding Complex
  Systems. Springer Berlin Heidelberg, 2012, 3--22.

\bibitem{rickart2012}
B\"{o}rner, K., Conlon, M., Corson-Rikert, J., and Ding, Y., Eds.
\newblock {\em VIVO: A Semantic Approach to Scholarly Networking and
  Discovery}.
\newblock Synthesis Lectures on the Semantic Web: Theory and Technology. Morgan
  and Claypool, 2012.

\bibitem{borner2012vivo}
B{\"o}rner, K., Conlon, M., Corson-Rikert, J., and Ding, Y.
\newblock Vivo: A semantic approach to scholarly networking and discovery.
\newblock {\em Synthesis Lectures on The Semantic Web: Theory and Technology
  7}, 1 (2012), 1--178.

\bibitem{borner2010multi}
B{\"o}rner, K., Contractor, N., Falk-Krzesinski, H., Fiore, S., Hall, K.,
  Keyton, J., Spring, B., Stokols, D., Trochim, W., and Uzzi, B.
\newblock A multi-level systems perspective for the science of team science.
\newblock {\em Science Translational Medicine 2}, 49 (2010), 49cm24.

\bibitem{Carayol2006}
Carayol, N., and Matt, M.
\newblock Individual and collective determinants of academic scientists'
  productivity.
\newblock {\em Information Economics and Policy\/} (2006).

\bibitem{conlon2009vivo}
Conlon, M.
\newblock {VIVO}: Enabling national networking of scientists.
\newblock
  \url{http://plaza.ufl.edu/mconlon/VIVO%20Overview%20OSTP%2020091112.pdf},
  Accessed February 1, 2013.

\bibitem{dijk2010}
Dijk, E.
\newblock Narcis, linking criss and oars in the netherlands: A matter of
  standards and identifiers., 2010.
\newblock Position paper presented at the Eurocris Workshop on CRIS, CERIF and
  institutional repositories, Rome, 10-11 May 2010.

\bibitem{dupuis2011}
Dupuis, M.~C.
\newblock Towards a joint current research information system for dutch higher
  education and research.
\newblock Online Newsletter Issue 23, Association for Learning Technology,
  2011.
\newblock \url{http://bit.ly/11dmbGo}.

\bibitem{Falagas2008}
Falagas, M., Ierodiakonou, V., and Alexiou, V.
\newblock At what age do biomedical scientists do their best work?
\newblock {\em Faseb Journal 22}, 12 (2006), 4067--4070.

\bibitem{DOLCE}
Gangemi, A., Guarino, N., Masolo, C., Oltramari, A., and Schneider, L.
\newblock Sweetening ontologies with dolce.
\newblock In {\em Proceedings of E-KAW} (2002).
\newblock \url{http://www.loa.istc.cnr.it/DOLCE.html}.

\bibitem{garfield1979citation}
Garfield, E.
\newblock {\em Citation indexing: Its theory and application in science,
  technology, and humanities}.
\newblock Wiley New York, 1979.

\bibitem{godin2005measurement}
Godin, B.
\newblock {\em Measurement and Statistics on Science and Technology: 1930 to
  the Present}, vol.~22.
\newblock Routledge, 2005.

\bibitem{BFO}
Grenon, P., and Smith, B.
\newblock Basic formal ontology ({BFO}), 2002.
\newblock \url{http://www.ifomis.org/bfo}.

\bibitem{Helbing:2012fk}
Helbing, D.
\newblock Accelerating scientific discovery by formulating grand scientific
  challenges.

\bibitem{LKIF}
Hoekstra, R., Breuker, J., {Di Bello}, M., and Boer, A.
\newblock The lkif core ontology of basic legal concepts.
\newblock In {\em Proceedings of the Workshop on Legal Ontologies and
  Artificial Intelligence Techniques (LOAIT)} (2007).

\bibitem{Horlings2012}
Horlings, E., and Gurney, T.
\newblock Search strategies along the academic lifecycle.
\newblock {\em Scientometrics\/} (2012).

\bibitem{huberman2001}
Huberman, B.
\newblock {\em {The laws of the Web}}.
\newblock MIT Press, 2001.

\bibitem{Levin1991}
Levin, S., and Stephan, P.
\newblock Research productivity over the life-cycle—evidence for academic
  scientists.
\newblock {\em American Economic Review 81}, 1 (1991), 114--132.

\bibitem{Lezcano2012}
Lezcano, L., J\"{o}rg, B., and Sicilia, M.
\newblock {Modeling the Context of Scientific Information: Mapping {VIVO} and
  {CERIF}}.
\newblock {\em Advanced Information Systems\/} (2012), 123--129.
\newblock \url{http://www.springerlink.com/index/Q5K7857961422017.pdf}.

\bibitem{SUMO}
Niles, I., and Pease, A.
\newblock Towards a standard upper ontology.
\newblock In {\em Proceedings of the 2nd International Conference on Formal
  Ontology in Information Systems (FOIS)} (2001).
\newblock \url{http://www.ontologyportal.org/}.

\bibitem{oskam2006harvex}
Oskam, M., Simons, H., and Mijnhardt, W.
\newblock Harvex: Integrating multiple academic information resources into a
  researcher's profiling tool.
\newblock In {\em Enabling Interaction and Quality: Beyond the Hanseatic League
  (8th International Conference on Current Research Information Systems)},
  A.~G. S. S. E.~J. Asserson, Ed., Leuven University Press (2006), 167--177.

\bibitem{powelletal05}
Powell, A., Nilsson, M., Naeve, A., and Johnston, P.
\newblock Dublin core metadata initiative - abstract model, 2005.
\newblock White Paper.

\bibitem{derek1963solla}
Price, D. d.~S.
\newblock {\em Little Science, Big Science}.
\newblock New York: Columbia University Press, 1963.

\bibitem{2013arXiv1301.5177R}
{Reijnhoudt}, L., {Costas}, R., {Noyons}, E., {Boerner}, K., and {Scharnhorst},
  A.
\newblock {''Seed+Expand'': A validated methodology for creating high quality
  publication oeuvres of individual researchers}.
\newblock {\em ArXiv e-prints\/} (Jan. 2013).

\bibitem{reijnhoudt2012}
Reijnhoudt, L., Stamper, M.~J., B{\"o}rner, Katy;~Baars, C., and Scharnhorst,
  A.
\newblock Narcis: Network of experts and knowledge organizations in the
  netherlands, 2012.
\newblock \url{http://cns.iu.edu/research/2012_NARCIS.pdf}, Accessed January
  26, 2013.

\bibitem{Scharnhorst2012}
Scharnhorst, A., B\"{o}rner, K., and Besselaar, P., Eds.
\newblock {\em {Models of Science Dynamics}}.
\newblock Understanding Complex Systems. Springer Berlin Heidelberg, Berlin,
  Heidelberg, 2012.

\bibitem{scharnhorst2010tracing}
Scharnhorst, A., and Garfield, E.
\newblock Tracing scientific influence.
\newblock {\em International Journal - Dynamics of Socio-Economic Systems 2}, 1
  (2010), 1--33.

\bibitem{scharnhorst2006webindicators}
Scharnhorst, A., and Wouters, P.
\newblock Webindicators: a new generation of {S}\&{T} indicators.
\newblock {\em Cybermetrics 10\/} (2006),
  http://cybermetrics.cindoc.csic.es/articles/v10i1p6.html.

\bibitem{sheppard2010learning}
Sheppard, N.
\newblock Learning how to play nicely: Repositories and cris, 2010.
\newblock \url{http://www.ariadne.ac.uk/issue64/wrn-repos-2010-05-rpt/}
  ,Accessed January 25, 2013.

\bibitem{DBLP:journals/tkde/ShvaikoE13}
Shvaiko, P., and Euzenat, J.
\newblock Ontology matching: State of the art and future challenges.
\newblock {\em IEEE Trans. Knowl. Data Eng. 25}, 1 (2013), 158--176.

\bibitem{most2012}
Van~der Most, F.
\newblock The role of evaluations in the development of researchers' careers. a
  conceptual frame and research strategy for a comparative study.
\newblock Poster presented at the conference `How to track researchers'
  careers.', Luxembourg, 9-10 February 2012. (unpublished, contact author),
  2012.

\bibitem{acumenproject}
Wouters, P.
\newblock Academic careers understood through measurements and norms.
\newblock \url{http://research-acumen.eu/}, Accessed January 26, 2013.

\bibitem{wouters2012users}
Wouters, P., and Costas, R.
\newblock Users, narcissism and control - tracking the impact of scholarly
  publications in the 21 st century.
\newblock \url{http://www. surffoundation.nl/nl/publicaties/Documents/Users\%
  20narcissism\%20and\%20control.pdf}, Accessed January 25, 2013.

\end{thebibliography}
\end{document}